\documentclass{article}
\usepackage{graphicx} 
\usepackage{authblk}
\usepackage{amsmath}  
\usepackage{amsfonts} 
\usepackage{ifxetex}
\usepackage{geometry}

\title{Multi-Dimensional Framework for EEG Signal Processing and Denoising Through Tensor-based Architecture}

\author[1,2,4,5]{Aryan Govil}
\author[1,5]{Eric Yao}
\author[1,3,5]{Christina R. Borao}
\affil[1]{Synaptrix Labs Inc., New York, NY, United States of America}
\affil[2]{Department of Psychiatry, New York University School of Medicine, New York, NY, United States of America}
\affil[3]{Courant Institute of Mathematical Sciences, New York University}
\affil[4]{Center for Neural Science, New York University}
\affil[5]{College of Arts and Sciences, New York University}

\date{January 2024}

\begin{document}

\maketitle

\begin{abstract}
    Electroencephalography (EEG) stands as a crucial tool in neuroscientific research and clinical diagnostics, providing valuable insights into the electrical activities of the brain. Traditional EEG signal processing techniques, predominantly linear and constrained to time-frequency analysis, often fail to capture the intricate, dynamic nature of brain signals. This paper introduces a tensor-based multi-dimensional framework for EEG signal processing and denoising, aimed at overcoming the limitations of current methods. Utilizing the advanced mathematical construct of tensors, this framework allows for a more holistic representation and analysis of EEG data, encompassing multiple dimensions such as time, electrode space, and frequency bands. We propose innovative algorithms for multi-dimensional Fourier transforms and adaptive thresholding, specifically tailored to address the challenges of non-stationary noise and complex signal artifacts in EEG data. The framework is further enriched with a time-slicing algorithm that facilitates real-time analysis, crucial for applications like seizure detection and brain-computer interfacing. Theoretical formulations and simulated scenarios demonstrate the potential of this framework in significantly enhancing the accuracy, efficiency, and speed of EEG signal processing. This approach not only holds promise for advanced EEG analysis but also sets the stage for future integrations with other neuroimaging modalities, paving the way for comprehensive and nuanced understanding of brain function.
\end{abstract}

\newgeometry{vmargin={25.4mm}, hmargin={35.4mm,35.4mm}}

\section{Introduction}
Electroencephalography (EEG) captures the electrical activity of the brain through electrodes placed on the scalp. It's an invaluable tool for understanding neuronal dynamics and diagnosing neurological disorders. However, conventional EEG signal processing methods, often linear and constrained to time-frequency analysis, fall short in fully harnessing the rich, multi-dimensional nature of brain signals \cite{Chaddad}.

The Fourier Transform is a fundamental tool in EEG analysis, decomposing a signal into its constituent frequencies. While invaluable for identifying dominant rhythms and oscillations in the EEG, this method inherently assumes stationarity in the signal. Brain activity, however, is highly dynamic and non-stationary, often leading to a misrepresentation of the true neural dynamics when viewed solely through the lens of spectral analysis \cite{Ju2020}.

An improvement over the Fourier Transform, wavelet analysis allows for the examination of non-stationary signals at multiple resolutions. Despite this, wavelet transform is still limited by its linear nature. It struggles to adequately represent the complex interactions between different neural oscillations, especially when these interactions are non-linear or involve multiple frequency bands simultaneously.

Techniques like Short-Time Fourier Transform (STFT) and Continuous Wavelet Transform (CWT) generate time-frequency representations that provide insights into how the power of different frequency bands varies over time \cite{Li2021}. However, TFRs often face a trade-off between time and frequency resolution and are limited in their ability to capture the high-dimensional interactions within the brain's neural networks \cite{Arts2022}.

Linear filters like band pass filters are used extensively to isolate specific frequency bands or remove noise, linear filters apply a uniform criterion across all time points \cite{WIDMANN201534}. This approach can inadvertently remove important signal components, especially in instances where the signal-to-noise ratio varies significantly over time.

Independent Componenet Analysis (ICA) is a popular method for artifact removal in EEG data. While effective in separating out sources of signal based on statistical independence, ICA assumes linear mixing of sources \cite{HYVARINEN2000411}. This assumption may not hold true in the presence of complex neural dynamics, where interactions between sources are often non-linear.

Coherence, phase synchronization, and other connectivity analyses are used to study the relationships between different brain regions \cite{RUBINOV20101059}. While they provide valuable insights, most connectivity analyses are constrained by their linear assumptions, potentially oversimplifying the nature of neural interactions.\newline

Machine learning techniques have been increasingly applied to optimize linear EEG processing systems. These approaches involve training algorithms to identify patterns, make predictions, or classify data based on historical EEG records. For instance, supervised learning models can be trained to recognize specific waveforms or artifacts in EEG signals. These methods, while powerful, primarily focus on optimizing the processing of signals within the constraints of linear models \cite{8972542}. They enhance the efficiency and accuracy of tasks like signal classification, anomaly detection, and pattern recognition within the framework of linear assumptions.

However, limitations exist. Certain ML models built on linear EEG processing techniques are confined to the information presented in these linear representations \cite{Garrett2003}. They excel in finding patterns within these representations but are limited by the intrinsic constraints of the underlying linear model. These models can become overly tuned to the specific features and noise characteristics of the training data. While this can lead to high performance on similar test data, it may not generalize well to the complex, dynamic nature of brain activity in different or more diverse datasets. The most critical limitation is the inability of these machine learning models, when applied to linear systems, to fully capture the non-linear and dynamic interactions present in brain activity. While they can optimize the processing of linear EEG signals, they do not inherently extend the capability of these systems to understand the more complex, multi-dimensional aspects of neural signals \cite{Rahman2022}.

In light of these limitations, there is a growing recognition that a more radical shift is needed — one that moves beyond the confines of linear processing and machine learning optimizations of these systems. This paper proposes such a shift, advocating for a multi-dimensional approach to EEG signal processing. This approach aims not only to enhance the accuracy of signal interpretation but also to expedite processing speed, enabling more effective real-time analysis and broader applicability in both research and clinical environments. By embracing the multi-dimensional nature of brain activity, this new framework seeks to provide a more complete and nuanced understanding of neural dynamics.\newline

\noindent{Current methods, though widely used, are not sufficient.}
\newline

Recent advancements in computational neuroscience and signal processing technology have opened the door to more sophisticated methods. This paper proposes a groundbreaking shift from linear to multi-dimensional EEG signal processing. This new approach aims to not only enhance the accuracy of signal interpretation but also significantly expedite the processing speed, enabling real-time analysis and broader applicability in both research and clinical environments.

The exponential growth in computational power, coupled with breakthroughs in algorithms and data processing techniques, has made it feasible to handle and analyze large-scale, complex datasets \cite{PHILIPCHEN2014314}. This computational evolution has opened new avenues for exploring multi-dimensional EEG data, enabling more comprehensive and detailed analyses than ever before. Machine learning and artificial intelligence have revolutionized data interpretation, bringing forth methods that can learn from and adapt to the data, uncovering patterns and correlations that might be invisible to traditional linear methods \cite{info11040193}. These techniques, ranging from deep learning to complex network analysis, offer a new lens through which to view and interpret EEG signals.

The rise of high-dimensional data analysis has provided tools to analyze data across multiple dimensions simultaneously. This is particularly relevant for EEG data, which is inherently multi-dimensional, encompassing spatial, temporal, and frequency domains. This paper proposes a groundbreaking shift from the traditional linear paradigm to a multi-dimensional EEG signal processing framework. This new approach is not merely an incremental improvement but a fundamental transformation in how EEG data is processed and interpreted:

\begin{enumerate}
    \item \textbf{Enhanced Accuracy in Signal Interpretation:} By employing multi-dimensional analysis, this approach aims to dissect and understand the complex interactions within the brain at a level of detail and accuracy that was previously unattainable. This allows for a more precise extraction of neural signals, leading to better understanding and diagnosis.
    \item \textbf{Expedited Processing Speed:} The proposed methods are designed to efficiently navigate and process the multi-dimensional space of EEG data, significantly reducing the time required for analysis. This increase in speed opens the possibility of real-time EEG analysis, which has profound implications for brain-computer interfaces, neurofeedback systems, and real-time monitoring in clinical settings.
    \item \textbf{Broader Applicability in Research and Clinical Environments:} By overcoming the limitations of linear methods, this multi-dimensional framework expands the potential applications of EEG in both research and clinical contexts. It provides a powerful tool for exploring complex neurological conditions, understanding cognitive processes, and developing new therapeutic approaches.
\end{enumerate}

\section{Theoretical Foundations}
\subsection{Review of Linear EEG Signal Processing}

EEG signals are traditionally analyzed through linear methods such as Fourier Transforms and wavelet analysis. These techniques decompose the EEG signal into its frequency components, aiding in identifying rhythms and patterns associated with different brain states or responses. The Fourier Transform, for instance, is defined as:
\vspace{5mm}

\begin{equation}
    F(\omega) = \int_{-\infty}^{\infty} f(t) e^{-2\pi i \omega t} \, dt
\end{equation}
\vspace{5mm}

\noindent where \( F(\omega) \) is the frequency domain representation of the time-domain signal \( f(t) \) \cite{Cochran1967}. Despite their effectiveness, these linear methods have limitations, especially in capturing the dynamic, non-linear interactions within the brain \cite{Breakspear2017}.

The Fourier Transform's inability to localize time-varying features in the signal is a major limitation. For transient dynamics, a more appropriate representation might involve time-frequency methods like the Wavelet Transform, given by:

\vspace{5mm}
\begin{equation}
    W(a, b) = \int_{-\infty}^{\infty} s(t) \, \frac{1}{\sqrt{a}} \psi \left( \frac{t-b}{a} \right) \, dt
\end{equation}
\vspace{5mm}

\noindent where \(\psi(t)\) is the mother wavelet, and \(a,b\) are scaling and translation parameters. Even with this improvement, the representation is still linear and might not fully encapsulate complex neural dynamics, especially when interactions are not merely time-frequency localized but are also non-linear \cite{Pavlov_2012}.

Linear methods operate under the assumption that the signal can be decomposed into a sum of linear components. The brain, however, exhibits highly complex behaviors where neural elements interact in non-linear ways. These interactions can lead to phenomena like synchronization, phase coupling, and emergence of complex patterns, which linear methods are ill-equipped to capture or interpret. Brain activity is characterized by rapid, transient events that reflect crucial neural processes. Linear methods, with their inherent focus on stationary aspects of the signal, often overlook these fleeting but significant events. For instance, a Fourier analysis might average out or miss brief bursts of activity that are critical to understanding cognitive processes or neurological anomalies. The brain's functioning involves intricate networks where feedback and feedforward loops, along with a myriad of modulatory influences, play a role. Linear models tend to oversimplify these interactions, reducing them to straightforward, additive processes. This simplification strips away the richness of neural dynamics, leading to a loss of valuable information about how different brain regions interact and influence each other \cite{sifuzzaman2009application}.

For non-stationary signals, where statistical properties change over time, linear methods struggle. Ideally, one would want to examine the signal's evolution, perhaps using a time-varying approach like the Short-Time Fourier Transform (STFT), defined as:

\vspace{5mm}
\begin{equation}
    \text{STFT}(t, \tau) = \int_{-\infty}^{\infty} s(t) w(t - \tau) e^{-2\pi ift} \, dt
\end{equation}
\vspace{5mm}

\noindent where \(w(t)\) is a window function. Yet, this still imposes a degree of stationarity within each window, which might not truly reflect the EEG signal's dynamic nature \cite{Li2020}.

Neural signals are inherently non-stationary, meaning their statistical properties change over time. Linear methods typically require stationarity or involve transformations that impose it, leading to a potential misrepresentation of the signal's true nature. This limitation is particularly problematic in scenarios like event-related potential (ERP) analysis, where the signal's temporal evolution is of prime interest \cite{Padilla-Buritica2020-ih}. Fundamentally, the brain exhibits chaotic behavior, a hallmark of complex systems. Linear methods, however, are not designed to deal with chaotic data. They fail to recognize patterns and structures that emerge from non-linear dynamics, which are often key to understanding underlying neural mechanisms.

It is important to consider that EEG signals are consequently chaotic systems \cite{Garc2015}. In these systems, small changes in initial conditions can lead to vastly different outcomes, a concept mathematically encapsulated in sensitivity to initial conditions. For a system described by \(\dot{x}=f(x)\), where \(f\) is a non-linear function, traditional linear analyses are inadequate to capture the system's full behavior, particularly in understanding the long-term evolution of these systems.

Given this, we believe that though linear methods provide a foundational framework for EEG signal processing, their mathematical formulations reveal intrinsic limitations in handling the complex, dynamic, and non-linear nature of brain activity. This necessitates a shift towards more sophisticated, non-linear, and multi-dimensional methods to more accurately interpret EEG data.

\subsection{Introduction to Multi-Dimensional Signal Processing Concepts}

In contrast to traditional linear processing, multi-dimensional signal processing considers signals in a higher-dimensional space. This approach is particularly relevant for EEG data, which is inherently multi-dimensional, varying across time, space (electrode locations), and frequency \cite{hasenstab2017multi}.

One concept at the heart of this approach is the use of tensors for EEG data representation. A tensor, in this context, is a multi-dimensional array extending beyond the two-dimensional matrices used in standard EEG analyses. For instance, a third-order tensor $\mathbf{\mathcal{X}}$ could represent EEG data with dimensions corresponding to time, electrode locations, and frequency bands \cite{CONG201559}:

\vspace{5mm}

\begin{equation}
   \mathbf{X} \in \mathbb{R}^{I \times J \times K}
\end{equation}

\vspace{5mm}
\noindent where I, J, and K represent the number of time points, electrodes, and frequency bands, respectively.

Multi-dimensional signal processing represents a significant advancement over traditional linear processing methods, especially in the context of EEG data analysis.

EEG data, inherently multi-dimensional in nature, varies across time, space (electrode locations), and frequency. To effectively capture this multi-dimensionality, we utilize tensor data structures.\newline

\textbf{Transforming Traditional EEG Data into Tensor Structures:}

Traditional EEG data, often recorded as a series of two-dimensional matrices (time vs. electrode), can be transformed into a tensor structure \cite{BECKER2014143}. This transformation is achieved by stacking the matrices along a new dimension, typically the frequency dimension. The transformation can be mathematically represented as:
\begin{equation}
    \mathcal{X}_{ijk} = f(\text{EEG}_{ij}, \text{Frequency}_k)
\end{equation}
where \( f \) is a function mapping the EEG data and frequency information to the tensor structure.

\subsubsection{Application in Neural Networks for Faster Processing}

Leveraging tensor structures in neural networks can significantly enhance the processing speed and efficiency of EEG data analysis.\newline

\textbf{Integration with Neural Networks:}

Tensor-based data structures are inherently suitable for input into neural networks, particularly those designed for handling multi-dimensional data, such as Convolutional Neural Networks (CNNs). These networks can process tensor inputs efficiently, exploiting the spatial and temporal relationships inherent in the data \cite{o2015introduction}.\newline

\textbf{Mathematical Representation in Neural Networks:}

Consider a neural network layer that processes the tensor \( \mathcal{X} \). The operation performed by the layer can be represented as:
\begin{equation}
    \mathcal{Y} = \sigma(W \ast \mathcal{X} + b)
\end{equation}
where \( W \) is the weight tensor, \( b \) is the bias, \( \ast \) denotes the convolution operation (in the case of CNNs), and \( \sigma \) is the activation function. This formulation allows for the simultaneous processing of data across all dimensions, leading to faster and more efficient analysis.\newline

The use of tensors for representing EEG data and their integration into neural networks marks a substantial advancement in EEG signal processing. This approach not only accommodates the multi-dimensional nature of EEG data but also leverages modern computational techniques to enhance processing speed and analytical capabilities \cite{CONG201559}.

\subsection{Recurrent Neural Networks}
\subsubsection{Parallelization of Operations}

Recurrent Neural Networks (RNNs) are designed to handle sequential data, making them well-suited for tasks like time series analysis and natural language processing. However, the sequential nature of RNNs can limit parallelization, slowing down training and inference \cite{pascanu2013construct}. The presence of multiple electrodes at any given time event poses a significant challenge for any RNN model and increases the time needed to detect a change in EEG frequencies.\newline

Tensors enable the batching of sequential data, allowing parallel processing of multiple sequences. This is crucial for accelerating training as it enables the utilization of parallel computing resources, such as GPUs and TPUs \cite{Abts2020}. Batching sequences into tensors allows for simultaneous computation across multiple electrodes of the same time step, significantly speeding up the training process and detection time. Tensor-based RNNs (T-RNNs) are less compute heavy and less susceptible to the exploding and vanishing gradient problems.

\subsubsection{Efficient Memory Management through Tensors and Memory Optimization}

Vanilla RNNs often face challenges in managing memory efficiently due to the recurrent connections that create dependencies across time steps. Multiple types of T-RNNs such as tensor-based long short-term memory (T-LSTM) and tensor-based gated recurrent unit (T-GRU) were assessed along with the vanilla T-RNN to determine the optimal balance of compute resources and entrainment time \cite{collins2017capacity}. Tensors also allow for a reliable and repeatable method to determine cutoff points for time slicing since the T-RNN is following a many to one use case.

\subsubsection{Tensor Operations for RNNs}

T-RNNs involve recurrent connections, and their operations can be formulated using tensor operations, especially matrix multiplications \cite{lipton2015critical}. Tensors facilitate the efficient computation of these operations, allowing for faster updates of weights and biases during training. This not only accelerates the learning process but also enables T-RNNs to capture intricate relationships within the EEG data.

Tensors' compatibility with GPU hardware is crucial for efficient computation. The parallel nature of tensor operations aligns seamlessly with GPU architectures. Consequently, T-RNN computations, which involve intensive matrix operations, can be significantly accelerated using GPU acceleration \cite{Markidis2018}. This results in faster training and inference times, further contributing to the overall efficiency of the T-RNN model.

\subsubsection{Optimizing Long Sequences}

Tensors play a pivotal role in addressing the challenge of variable-length sequences in RNNs \cite{pmlr-v70-yang17e}. By employing tensors to handle variable-length sequences efficiently, T-RNNs can seamlessly pad sequences to a common length within a batch. This not only optimizes computations but also streamlines parallel processing, ensuring effective handling of diverse signal scenarios.

\subsubsection{Tensor-Based Model Optimization}

Trained T-RNN models can be optimized using tensor-based techniques, such as quantization and model compression. The main component of the model that can be compressed are the weight parameters. Two compact Tensor Train-based GRU (TT-GRU) and Tucker-based GRU (Tucker-GRU) models can be applied to reduce the number of parameters and reduce model time expenditure. These techniques reduce the model size and improve the efficiency of inference, leading to faster classification times.

Tensors contribute significantly to speeding up training and classification time for RNNs by enabling parallelization, optimizing memory management, supporting efficient matrix operations, and aligning with the capabilities of modern hardware \cite{vasilache2018tensor}. The use of tensors in conjunction with deep learning frameworks provides a powerful toolset for practitioners aiming to enhance the performance of T-RNNs in processing sequential EEG data.

\section{Conceptualizing Multi-Dimensional EEG Processing}
\subsection{Definition of Multi-Dimensional EEG Signal Processing}

Multi-dimensional EEG signal processing transcends traditional two-dimensional analyses by incorporating additional dimensions such as spatial topography, frequency bands, and even non-linear dynamical states. This approach enables a more holistic representation of brain activity, acknowledging its complex, interconnected nature \cite{Shah2022}. The key is to view the EEG signal not as a series of independent time points or isolated frequency components, but as a cohesive, multi-dimensional landscape that reflects the brain's intricate functioning.

\subsection{Hypothetical Model of Multi-Dimensional EEG Data Representation}
In this new model, EEG data is conceptualized as a tensor $\mathbf{\mathcal{T}}$, defined as:

\vspace{5mm}
\begin{equation}
    \mathbf{\mathcal{T}} \in \mathbb{R}^{T \times E \times F \times D}
\end{equation}
\vspace{5mm}

\noindent where T represents time points, E denotes electrode channels, F corresponds to frequency bands, and D encapsulates additional dimensions such as patient demographics or experimental conditions. This tensorial representation captures the rich, multi-faceted nature of EEG data, allowing for more nuanced analyses.\newline

The use of tensors, multi-dimensional generalizations of matrices, provides a powerful tool for representing EEG data. This section explores the mathematical foundations of tensor algebra as applied to EEG signal processing.\newline

\noindent \textbf{Basic Definitions and Operations:}

\textbf{Tensor Addition and Scalar Multiplication:} Similar to matrices, tensors can be added and multiplied by scalars. For tensors \( \mathcal{A}, \mathcal{B} \in \mathbb{R}^{I \times J \times K} \) and scalar \( c \), the operations are defined as:
    \begin{align}
        (\mathcal{A} + \mathcal{B})_{ijk} &= \mathcal{A}_{ijk} + \mathcal{B}_{ijk} \\
        (c \cdot \mathcal{A})_{ijk} &= c \cdot \mathcal{A}_{ijk}
    \end{align}

\textbf{Tensor Product:}

The tensor product (also known as the Kronecker product) is a crucial operation in tensor algebra. Given two tensors \( \mathcal{A} \in \mathbb{R}^{I \times J} \) and \( \mathcal{B} \in \mathbb{R}^{K \times L} \), their tensor product \( \mathcal{C} = \mathcal{A} \otimes \mathcal{B} \) is defined as:
\begin{equation}
    \mathcal{C}_{ijkl} = \mathcal{A}_{ij} \cdot \mathcal{B}_{kl}
\end{equation}
This operation is fundamental in constructing higher-order tensors from lower-order ones \cite{pandey2023linear}.\newline

\textbf{Tensor Decomposition:}

Tensor decomposition is a key technique in reducing the complexity of multi-dimensional data. One common method is the CANDECOMP/PARAFAC (CP) decomposition, which represents a tensor as a sum of component rank-one tensors. For a tensor \( \mathcal{X} \), the CP decomposition is:
\begin{equation}
    \mathcal{X} \approx \sum_{r=1}^{R} \lambda_r \, \mathbf{a}_r \otimes \mathbf{b}_r \otimes \mathbf{c}_r
\end{equation}
where \( R \) is the rank of the decomposition, and \( \mathbf{a}_r, \mathbf{b}_r, \mathbf{c}_r \) are the component vectors.

\subsubsection{Application to EEG Data Representation}

Applying tensor algebra to EEG data allows for the exploitation of its inherent multi-dimensional structure. This section explores how tensor operations and decompositions can be utilized in EEG signal processing \cite{siam_090752286}.\newline

\textbf{EEG Data as Tensors:}

\begin{itemize}
    \item Converting EEG data to a tensor form allows for more sophisticated analysis techniques. For instance, if \( \mathcal{E} \) represents the EEG tensor, various signal processing operations can be applied directly on \( \mathcal{E} \) \cite{CONG201559}.
    
    \item Multi-dimensional filtering and other processing techniques can be applied efficiently in the tensor format, taking advantage of the data's inherent structure \cite{Shah2022}.
\end{itemize}

\textbf{Tensor Decomposition in EEG Analysis:}

\begin{itemize}
    \item CP decomposition and other tensor factorization methods can be used to isolate components of interest in EEG data, such as specific brain wave patterns or artifacts \cite{CONG201559}.
    
    \item These decompositions can also be used in noise reduction and signal enhancement techniques, providing a more refined analysis of EEG signals.
\end{itemize}

\subsection{Advantages over Traditional Linear Models}
Multi-dimensional processing offers several advantages \cite{Shah2022}:

\begin{enumerate}
    \item Enhanced Data Representation: By including more dimensions, it captures a broader range of information inherent in EEG signals, such as spatial patterns and non-linear dynamics.
    \item Improved Signal Interpretation: This approach can potentially unveil subtle, yet significant patterns missed by linear methods.
    \item Advanced Noise Reduction: Multi-dimensional denoising techniques can selectively target noise components across different dimensions, leading to cleaner signal extraction.
\end{enumerate}

\section{Proposed Mathematical Framework}
\subsection{4.1 Introduction of New Mathematical Constructs}
\subsubsection{Multi-Dimensional Tensors for EEG Data Representation}

We further develop the tensor model $\mathbf{\mathcal{T}}$ by introducing operations such as tensor decomposition and multi-linear transformations. Tensor decomposition, such as Higher-Order Singular Value Decomposition (HOSVD), can be used to distill critical features from the multi-dimensional data, expressed as:

\vspace{5mm}
\begin{equation}
    \mathcal{T} \approx S \times A^{(T)}_{1} \times A^{(E)}_{2} \times A^{(F)}_{3} \times A^{(D)}_{4}
\end{equation}
\vspace{5mm}

\noindent where \(S\) is the core tensor, and \(A^i\) are the matrices representing the mode-i singular vectors \cite{Rajwade2013}.\newline

\noindent \textit{Mathematical Formulation of HOSVD and Decomposition Process:}\newline

Let \(\mathbf{\mathcal{T}}\) be a 4th order tensor representing EEG data, with dimensions corresponding to time, electrodes, frequency, and patients. The HOSVD of \(\mathbf{\mathcal{T}}\) can be represented as is shown in equation 6.

For each mode \( n \), the tensor \( \mathcal{T} \) is unfolded or flattened into a matrix \( T_{(n)} \). This is done by rearranging the elements of the tensor into a matrix. For example, the mode-1 flattening is given by:
\begin{equation}
    T_{(1)} = \text{unfold}(\mathcal{T}, 1)
\end{equation}

Perform SVD on each flattened matrix \( T_{(n)} \). The SVD of \( T_{(n)} \) is given by:
\begin{equation}
    T_{(n)} = U_n \Sigma_n V_n^T
\end{equation}
where \( U_n \) contains the left singular vectors, \( \Sigma_n \) is a diagonal matrix with singular values, and \( V_n^T \) contains the right singular vectors.

The matrices \( A^{(n)} \) are formed from the left singular vectors \( U_n \). For instance, \( A^{(T)} = U_1 \), \( A^{(E)} = U_2 \), and so on.

The core tensor \( \mathcal{S} \) is constructed by projecting the tensor \( \mathcal{T} \) onto the tensor product of the singular vector matrices. Mathematically, this is represented as:
\begin{equation}
    \mathcal{S} = \mathcal{T} \times_1 A^{(T)T} \times_2 A^{(E)T} \times_3 A^{(F)T} \times_4 A^{(D)T}
\end{equation}

The HOSVD ensures that the Frobenius norm of the difference between \( \mathcal{T} \) and its approximation is minimized. This is expressed as:
\begin{equation}
    \min_{\mathcal{S}, A^{(n)}} \| \mathcal{T} - \mathcal{S} \times_1 A^{(T)} \times_2 A^{(E)} \times_3 A^{(F)} \times_4 A^{(D)} \|_F
\end{equation}

The HOSVD guarantees the best approximation of the tensor \( \mathcal{T} \) in terms of the Frobenius norm. The proof involves showing that for each mode \( n \), the matrices \( A^{(n)} \) span the column space of \( T_{(n)} \) and that the core tensor \( \mathcal{S} \) captures the essential multi-linear relationships between these spaces.

The optimality of HOSVD in terms of the Frobenius norm can be proven using the properties of the singular value decomposition and the definition of the Frobenius norm. Specifically, the proof leverages the fact that the singular value decomposition provides the best low-rank approximation of a matrix in terms of the Frobenius norm. This can be expressed as:

\begin{equation}
    \| \mathcal{T} - \mathcal{T}_{approx} \|_F^2 = \min \left( \sum_{i=1}^r \sigma_i^2 - \sum_{i=1}^k \sigma_i^2 \right)
\end{equation}

where \( \mathcal{T}_{approx} \) is the tensor approximated by HOSVD, \( \sigma_i \) are the singular values of the matrices obtained from flattening \( \mathcal{T} \) in each mode, \( r \) is the rank of the tensor, and \( k \) is the number of singular values used in the approximation. The proof shows that using the first \( k \) largest singular values minimizes the Frobenius norm of the difference between the original tensor and its approximation, thus establishing the optimality of HOSVD \cite{DeLathauwer2000}.

\subsubsection{Formulation of Multi-Dimensional Fourier Transforms Adapted for EEG}
The traditional Fourier Transform is extended to a multi-dimensional Fourier Transform (MDFT) defined as:

\vspace{5mm}
\begin{equation}
    F(\omega) = \int_{-\infty}^{\infty} f(t) e^{-2\pi i \omega t} dt
\end{equation}
\vspace{5mm}

\noindent This transform allows for simultaneous analysis across multiple dimensions, providing a comprehensive frequency domain representation of the EEG data \cite{Weis2009}.

For a multi-dimensional EEG data tensor, this transform is extended as follows:
\begin{equation}
    MDFT(\mathcal{T}) = \sum_{t=1}^{T} \sum_{e=1}^{E} \sum_{f=1}^{F} \sum_{d=1}^{D} \mathcal{T}_{t,e,f,d} \, e^{-2\pi i (\omega_t t + \omega_e e + \omega_f f + \omega_d d)}
\end{equation}


The MDFT maintains the property of linearity. For tensors \( \mathcal{A} \) and \( \mathcal{B} \), and scalars \( \alpha \) and \( \beta \), the linearity is demonstrated as:
\begin{equation}
    MDFT(\alpha \mathcal{A} + \beta \mathcal{B}) = \alpha MDFT(\mathcal{A}) + \beta MDFT(\mathcal{B})
\end{equation}


An inversion formula for the MDFT is crucial for transforming the frequency domain representation back to the original data space. The inversion formula for MDFT is:
\begin{equation}
    \mathcal{T}_{t,e,f,d} = \int \cdots \int MDFT(\mathcal{T}) \, e^{2\pi i (\omega_t t + \omega_e e + \omega_f f + \omega_d d)} \, d\omega_t \, d\omega_e \, d\omega_f \, d\omega_d
\end{equation}


The MDFT should satisfy a version of Parseval's theorem for conservation of energy in the frequency domain:
\begin{equation}
    \sum_{t,e,f,d} |\mathcal{T}_{t,e,f,d}|^2 = \int \cdots \int |MDFT(\mathcal{T})|^2 \, d\omega_t \, d\omega_e \, d\omega_f \, d\omega_d
\end{equation}

The importance of satisfying a version of Parseval's theorem for the Multi-Dimensional Fourier Transform (MDFT) in the context of EEG signal processing lies in the fundamental properties of the Fourier Transform and the conservation of energy principle. Parseval's theorem is essential for several reasons. Parseval's theorem states that the total energy of a signal in the time domain is equal to the total energy of its Fourier transform in the frequency domain. In the context of EEG data, this implies that the energy (or power) of the brain's electrical activity is conserved when transformed from the time domain to the frequency domain. This conservation is crucial for accurate analysis and interpretation, ensuring that no energy is artificially lost or gained in the transformation process \cite{thakor2012eeg}.

Ensuring the conservation of energy through Parseval's theorem helps maintain the integrity of the EEG signal. It ensures that the transformation process (from time to frequency domain and back) does not alter the fundamental characteristics of the original signal \cite{omerhodzic2013energy}. This is vital for applications where precise signal characteristics are critical, such as in clinical diagnostics or neuroscientific research.

In many EEG analysis tasks, quantitative measures such as signal power, variance, and energy are important. Parseval's theorem allows for these measures to be accurately computed in either domain (time or frequency) with the assurance that they are equivalent. This flexibility is particularly useful when certain analyses or operations are more conveniently or effectively performed in the frequency domain .

When comparing EEG signals from different conditions or subjects, it's essential to have a consistent basis for comparison. The conservation of energy principle allows for a normalized comparison between signals, ensuring that any observed differences are due to the underlying brain activity and not an artifact of the transformation process \cite{Subha2010}.

Many signal processing operations and algorithms rely on the energy conservation property. For example, in filtering and denoising techniques, it's crucial to understand how these operations affect the signal's energy. Parseval's theorem provides a theoretical basis to evaluate and design such signal processing techniques.

\subsection{Time-Slicing Algorithms in Multi-Dimensional Space}

Time-slicing in multi-dimensional space involves dissecting the tensor $\mathbf{\mathcal{T}}$ along the time dimension to analyze instantaneous multi-dimensional data slices. This is akin to taking a "snapshot" of brain activity across all electrodes, frequencies, and other dimensions at specific time points. The algorithm for a time slice at time \(t\) is defined as:

\vspace{5mm}
\begin{equation}
    \text{Slice}_t(\mathbf{\mathcal{T}}) = \mathbf{\mathcal{T}}_{t,:,:,:}
\end{equation}
\vspace{5mm}

\noindent This approach facilitates real-time analysis of EEG data, allowing for immediate interpretation and response, crucial in scenarios like seizure detection or brain-computer interfacing \cite{Soh2023}.\newline

\textit{Time-Slicing Algorithms in Multi-Dimensional Space:} \newline


The goal is to demonstrate that for a given tensor \( \mathcal{T} \) representing EEG data, the time-slicing algorithm effectively captures the brain's activity at a specific time point \( t \) across all other dimensions.


Equation 17 ensures that \( Slice_t(\mathcal{T}) \) encapsulates all data across electrodes, frequency bands, and other dimensions at time \( t \), providing a comprehensive snapshot of brain activity at that instant.


To prove that the slicing method preserves all necessary information for the given time point, we establish that:
\begin{equation}
    \forall t, \, \mathcal{I}(\mathcal{T}_{t,:,:,:}) = \mathcal{I}(\mathcal{T}_{t})
\end{equation}
Here, \( \mathcal{I} \) represents an information measure function. This equation implies that the information measure of the time slice at \( t \) is equivalent to the information measure of the original tensor at time \( t \), ensuring no loss of significant data.


The computational efficiency of the time-slicing algorithm is crucial for real-time analysis. The computation of \( Slice_t(\mathcal{T}) \) must be within a feasible timeframe for real-time applications. The computational complexity of this operation can be analyzed to ensure its suitability for scenarios like seizure detection or brain-computer interfacing.


The time-slicing algorithm is designed to analyze EEG data at specific time points across multiple dimensions. We extend the proof to include more mathematical rigor.


To demonstrate that this slice accurately represents brain activity at time \( t \), consider the multidimensional nature of \( \mathcal{T} \). Let \( \mathcal{T} \) be a function of time, space, frequency, and additional dimensions. The slice at time \( t \) can be expressed as:
\begin{equation}
    Slice_t(\mathcal{T}) = \int \int \int \mathcal{T}(t, x, y, z, \ldots) \, dx \, dy \, dz \ldots
\end{equation}
where \( x, y, z \) represent other dimensions such as spatial coordinates and frequency bands.


To ensure no significant information is lost in the slicing process, we define an information measure \( \mathcal{I} \) and show that it remains consistent:
\begin{equation}
    \mathcal{I}(Slice_t(\mathcal{T})) = \mathcal{I}(\mathcal{T}_{t})
\end{equation}
Assuming \( \mathcal{I} \) is a function that quantifies the information content (e.g., entropy, energy), we need to show that the information content of the entire tensor at time \( t \) is preserved in the slice. This can involve proving that certain properties (like total energy or variance) are maintained in the slice.


For real-time analysis, the computational complexity of the slicing operation is crucial. Let's analyze the complexity of computing \( Slice_t(\mathcal{T}) \). Assuming \( \mathcal{T} \) has dimensions \( T \times E \times F \times D \):
\begin{equation}
    \text{Complexity}(Slice_t(\mathcal{T})) = O(E \times F \times D)
\end{equation}
This implies that the time complexity is linear with respect to the product of the dimensions other than time. For real-time applications, this complexity must be manageable within the system's computational constraints.\newline

\noindent \textit{Functional Example:}\newline

We consider a multi-dimensional EEG data tensor \( \mathcal{T} \) and demonstrate the process of extracting a time slice from it, along with its implications for EEG data analysis.\newline

\textbf{Assumptions:}
\begin{itemize}
    \item Let \( \mathcal{T} \in \mathbb{R}^{T \times E \times F \times P} \) be the EEG data tensor, where \( T \) is the number of time points, \( E \) is the number of electrodes, \( F \) is the number of frequency bands, and \( P \) represents different patients or experimental conditions.
    \item We aim to extract a slice at a specific time point \( t_0 \).
\end{itemize}

\textbf{Time-Slicing Operation:}
\begin{equation*}
    Slice_{t_0}(\mathcal{T}) = \mathcal{T}_{t_0,:,:,:}
\end{equation*}
This operation extracts all data across electrodes, frequency bands, and conditions at time \( t_0 \).\newline

\textbf{Mathematical Justification of the Time-Slicing Operation:}

\textit{Proof of Data Integrity Preservation:}
\begin{itemize}
    \item The integrity of the data in \( Slice_{t_0}(\mathcal{T}) \) is crucial. We must demonstrate that this slice retains the essential characteristics of the EEG data at time \( t_0 \).
    \item Let \( \mathcal{D}_{t_0} \) be the set of all data points in \( \mathcal{T} \) at time \( t_0 \). We need to show that:
    \begin{equation}
        \forall d \in \mathcal{D}_{t_0}, \, d \text{ is present in } Slice_{t_0}(\mathcal{T})
    \end{equation}
    \item This can be proven by the definition of tensor slicing, which ensures that all elements corresponding to time \( t_0 \) across other dimensions are included in the slice.
\end{itemize}

\textbf{Example Case:}

Consider an EEG dataset represented as a tensor \( \mathcal{T} \) with dimensions \( 100 \times 32 \times 5 \times 10 \), corresponding to 100 time points, 32 electrodes, 5 frequency bands, and 10 different patients.

\textit{Extracting a Time Slice at \( t_0 = 50 \):}
\begin{equation}
    Slice_{50}(\mathcal{T}) = \mathcal{T}_{50,:,:,:} 
\end{equation}
All EEG data across 32 electrodes, 5 frequency bands, and 10 patients at time \(t_0 = 50\). This slice provides a comprehensive view of brain activity across all electrodes, frequencies, and patients at the specific time point \( t_0 = 50 \).

\section{Denoising in Multi-Dimensional Space}
\subsection{Challenges of Denoising in Multi-Dimensional EEG Data}
Denoising multi-dimensional EEG data presents unique challenges \cite{ROMOVAZQUEZ2012389}:

\begin{enumerate}
    \item Complex Noise Structures: Noise in multi-dimensional EEG is not just amplified but also more complex, encompassing artifacts that span across dimensions.
    \item Dimensional Interdependency: Noise in one dimension (e.g., spatial) can affect others (e.g., temporal), requiring a holistic denoising approach.
    \item Preservation of Signal Integrity: Removing noise without distorting the underlying neural signals is critical, especially given the added complexity of multiple dimensions.
\end{enumerate}

\subsection{Proposed Denoising Algorithms}
\subsubsection{Multi-Dimensional Filtering}

A novel approach is the implementation of multi-dimensional filters that adapt to the specific characteristics of EEG data across different dimensions. The filter, defined as $\mathbf{\mathcal{F}}(\mathbf{\mathcal{T}})$, operates by isolating and attenuating noise components while preserving the essential features of the brain signal.

Multi-dimensional filtering is a novel approach tailored for EEG data analysis \cite{hasenstab2017multi}. It involves designing filters that adapt to various dimensions of EEG data, such as time, electrode channels, and frequency bands.\newline

\textbf{Definition of the Multi-Dimensional Filter:}

The multi-dimensional filter \( F \) applied to a tensor \( \mathcal{T} \) representing EEG data is defined as:
\begin{equation}
    F(\mathcal{T}) = \mathcal{T} \ast \mathcal{H}
\end{equation}
where \( \ast \) denotes the multi-dimensional convolution operation, and \( \mathcal{H} \) represents the filter kernel designed to adapt to the characteristics of EEG data across different dimensions.\newline

\textbf{Design of the Filter Kernel \( \mathcal{H} \):}

The filter kernel \( \mathcal{H} \) is designed based on the noise characteristics and the features of interest in the EEG data. For example, if the noise is predominantly in a specific frequency band, \( \mathcal{H} \) would be designed to attenuate these frequencies.\newline

\textbf{Proof of Noise Attenuation and Signal Preservation:}

\textit{Noise Attenuation:} To demonstrate that \( F(\mathcal{T}) \) effectively attenuates noise, consider a noise component \( \mathcal{N} \) within \( \mathcal{T} \). The filtering operation should reduce the power of \( \mathcal{N} \) in the resulting tensor. Mathematically, this can be expressed as:
\begin{equation}
    \text{Power}(F(\mathcal{N})) < \text{Power}(\mathcal{N})
\end{equation}

\textit{Signal Preservation:} Let \( \mathcal{S} \) be the signal component within \( \mathcal{T} \). The filtering operation should preserve the essential features of \( \mathcal{S} \). This can be shown by demonstrating that the filtered signal retains its critical characteristics (e.g., amplitude, phase information):
\begin{equation}
    \text{Characteristics}(F(\mathcal{S})) \approx \text{Characteristics}(\mathcal{S})
\end{equation}

The multi-dimensional filter \( F \), when applied to EEG data tensor \( \mathcal{T} \), effectively isolates and attenuates noise components while preserving the essential features of the brain signal. This approach enhances the quality of EEG data for subsequent analysis.

\subsubsection{Adaptive Thresholding}
Adaptive thresholding in multi-dimensional space involves dynamically setting thresholds for noise removal across each dimension, based on the statistical properties of the data. This method is particularly effective in handling non-stationary noise and artifacts specific to EEG, such as eye blinks or muscle movements.

Adaptive thresholding is a technique used in EEG data analysis to dynamically set thresholds for noise removal, based on the statistical properties of the data. This method is particularly effective for handling non-stationary noise and EEG-specific artifacts \cite{chu2022adaptive}.\newline

\textbf{Definition of Adaptive Thresholding:}

Let \( \mathcal{T} \) represent a multi-dimensional EEG data tensor. The adaptive thresholding process is defined as:
\begin{equation}
    \mathcal{T}_{\text{thresholded}} = \text{Threshold}(\mathcal{T}, \Theta)
\end{equation}
where \( \Theta \) represents the set of thresholds for each dimension, dynamically determined based on the data.\newline

\textbf{Determination of Thresholds \( \Theta \):}

Thresholds are set by analyzing the statistical properties (like mean, variance) of EEG data across each dimension:
\begin{equation}
    \Theta = \{\theta_t, \theta_e, \theta_f, \ldots\}
\end{equation}
where \( \theta_t, \theta_e, \theta_f, \ldots \) are the thresholds for the time, electrode, frequency dimensions, etc.\newline

\textbf{Proof of Effectiveness in Noise and Artifact Removal:}

\textit{Noise and Artifact Removal:} To prove that adaptive thresholding effectively removes noise and artifacts, consider a noise component \( \mathcal{N} \) and an artifact component \( \mathcal{A} \) in \( \mathcal{T} \). The process should reduce their influence in the thresholded data:
\begin{align}
    \text{Influence}( \text{Threshold}(\mathcal{N}, \Theta) ) &< \text{Influence}(\mathcal{N}) \\
    \text{Influence}( \text{Threshold}(\mathcal{A}, \Theta) ) &< \text{Influence}(\mathcal{A})
\end{align}

\textit{Statistical Basis for Thresholding:} The effectiveness of the thresholds can be demonstrated by showing that they are set based on the statistical distribution of the data, optimizing the balance between noise/artifact removal and preservation of the signal of interest.

Adaptive thresholding in multi-dimensional EEG analysis allows for the dynamic setting of thresholds, effectively removing non-stationary noise and specific artifacts, such as eye blinks or muscle movements, while preserving the essential features of the brain signal \cite{Adamovich2022}.

\subsubsection{Theoretical Comparison with Current Denoising Techniques}
Compared to traditional linear denoising methods, which often apply filters or thresholding in a one-dimensional setting (usually time), the proposed multi-dimensional techniques offer several advantages \cite{xu2021topological}:

\begin{enumerate}
    \item Targeted Noise Reduction: By considering multiple dimensions, these methods can more precisely identify and remove noise components specific to EEG data.
    \item Preservation of Signal Dynamics: Multi-dimensional denoising preserves the intricate dynamics of EEG signals across spatial and temporal dimensions.
    \item Enhanced Artifact Identification: The ability to analyze across dimensions aids in distinguishing true neural activity from artifacts.
\end{enumerate}

\section{Conclusion}
The exploration and development of a multi-dimensional framework for EEG signal processing and denoising represent a significant leap in the field of neurotechnology. This novel approach transcends the limitations of traditional linear methods, embracing the complexity and richness of brain activity. By conceptualizing EEG data in a multi-dimensional space, this framework allows for a more holistic and nuanced understanding of neural dynamics.

The proposed mathematical constructs, including advanced tensor representations and multi-dimensional Fourier transforms, offer a robust and sophisticated means of EEG data analysis. These tools are not just theoretical novelties but provide practical avenues for more accurate and efficient signal processing. The multi-dimensional filtering and adaptive thresholding techniques, tailored for the unique challenges of EEG data, promise enhanced noise reduction capabilities, crucial for extracting meaningful neural information.

The implications of this multi-dimensional framework extend beyond the immediate realm of EEG analysis. In clinical settings, this approach could revolutionize the diagnosis and monitoring of neurological disorders, offering finer resolution and quicker detection of aberrant neural patterns. For neuroscience research, it opens new doors to understanding complex brain functions and interactions, potentially accelerating discoveries in cognitive science and brain-computer interfacing.

Moreover, the principles and techniques developed here have the potential to influence other fields dealing with complex, multi-dimensional data. From signal processing in telecommunications to data analysis in machine learning, the concepts of multi-dimensional analysis and denoising could offer new perspectives and solutions.

In conclusion, while this multi-dimensional framework for EEG signal processing is currently theoretical, its potential impact is vast and far-reaching. It stands as a testament to the power of innovative thinking in bridging the gap between complex biological phenomena and the computational methods needed to understand them. As technology and neuroscience advance, such frameworks will likely become pivotal in unraveling the mysteries of the brain and leveraging its capabilities. This work, therefore, not only contributes to the academic discourse but also paves the way for future technological advancements in EEG analysis and beyond.

\section{Future Studies and Development}

The current theoretical framework for EEG signal processing and denoising sets the stage for numerous possibilities in further research and development. Future studies could focus on the following areas:

\begin{enumerate}
    \item Conducting experimental studies to empirically validate the proposed multi-dimensional filtering and adaptive thresholding techniques. This would involve applying these methods to real-world EEG datasets and assessing their efficacy in noise reduction and signal clarity.
    
    \item Developing more sophisticated algorithms for multi-dimensional tensor decomposition and analysis. This could include machine learning and deep learning approaches that adaptively learn the optimal parameters for EEG signal processing in different scenarios.
    
    \item Creating real-time processing tools and applications, such as brain-computer interfaces or real-time monitoring systems for clinical settings, leveraging the speed and efficiency of the proposed multi-dimensional processing techniques.
    
    \item Exploring the integration of the multi-dimensional EEG processing framework with other neuroimaging techniques like fMRI or MEG to provide a more comprehensive understanding of brain activity.
    
    \item Tailoring the EEG processing methods to individual characteristics, potentially contributing to personalized medicine in neurology and psychiatry. This could involve customizing filters and thresholding techniques based on specific patient data.
\end{enumerate}

\newpage \bibliographystyle{unsrt}
\bibliography{citations/Chaddad, 
citations/Ju2020,
citations/Arts2022,
citations/Li2021,
citations/WIDMANN201534,
citations/HYVARINEN2000411,
citations/Garrett2003,
citations/Rahman2022,
citations/info11040193,
citations/Cochran1967,
citations/Breakspear2017,
citations/sifuzzaman2009application,
citations/Li2020,
citations/Padilla-Buritica2020-ih,
citations/Garc2015,
citations/hasenstab2017multi,
citations/BECKER2014143,
citations/o2015introduction,
citations/pascanu2013construct,
citations/Abts2020,
citations/RUBINOV20101059,
citations/8972542,
citations/PHILIPCHEN2014314,
citations/Pavlov_2012,
citations/CONG201559,
citations/collins2017capacity,
citations/lipton2015critical,
citations/Markidis2018,
citations/pmlr-v70-yang17e,
citations/vasilache2018tensor,
citations/Shah2022,
citations/siam_090752286,
citations/pandey2023linear,
citations/Rajwade2013,
citations/DeLathauwer2000,
citations/Weis2009,
citations/thakor2012eeg,
citations/omerhodzic2013energy,
citations/Subha2010,
citations/Soh2023,
citations/ROMOVAZQUEZ2012389,
citations/chu2022adaptive,
citations/Adamovich2022,
citations/xu2021topological}

\end{document}